\documentclass{article}

\PassOptionsToPackage{numbers,compress}{natbib}
\usepackage[preprint]{neurips_2026}

\usepackage[utf8]{inputenc}
\usepackage[T1]{fontenc}
\usepackage{graphicx}
\usepackage{hyperref}
\usepackage{url}
\usepackage{booktabs}
\usepackage{amsfonts}
\usepackage{amssymb}
\usepackage{microtype}
\usepackage{xcolor}
\usepackage{makecell}
\usepackage{bm}
\usepackage{amsmath}
\usepackage{enumitem}
\usepackage{etoolbox}
\usepackage{multirow}
\usepackage{color,colortbl}
\usepackage{tikz}
\usepackage{pgfplots}
\pgfplotsset{compat=1.18}
\usepgfplotslibrary{groupplots}
\usepackage{float}
\usepackage{wrapfig}
\usepackage{titlesec}
\usepackage[most]{tcolorbox}
\usepackage[noend]{algpseudocode}
\usepackage[ruled,linesnumbered,vlined]{algorithm2e}

\definecolor{mylightblue}{rgb}{0.5, 0.7, 1}
\newcommand{\ourDataset}{\textsc{HeterQA}}

\newtcolorbox{AIbox}[2][]{%
  colback=gray!5!white,
  colframe=gray!75!black,
  title={#2},
  fonttitle=\bfseries,
  arc=0.5mm,
  boxrule=0.8pt,
  left=2pt,
  right=2pt,
  top=2pt,
  bottom=2pt,
  #1
}

\appto{\ttfamily}{\hyphenchar\font=`\-}
\algrenewcommand\algorithmicforall{\textbf{for each}}

\titlespacing*{\subsection}{0pt}{2.5pt plus 0.5pt minus 0.5pt}{2.5pt plus 0.5pt minus 0.5pt}

\title{\ourDataset: Benchmarking Record Retrieval over Multiple Heterogeneous Sources}

\author{%
  \begin{tabular}{c}
  Yaodong Su$^{1}$ \quad Hanchang Li$^{1}$ \quad Quanqing Xu$^{2}$ \quad
  Chuanhui Yang$^{2}$ \quad Yixiang Fang$^{1}$ \\
  $^{1}$CUHK-Shenzhen \quad $^{2}$OceanBase, Ant Group \\
  \texttt{yaodongsu@link.cuhk.edu.cn} \quad
  \texttt{hanchangli@link.cuhk.edu.cn} \\
  \texttt{xuquanqing.xqq@oceanbase.com} \quad
  \texttt{rizhao.ych@oceanbase.com} \\
  \texttt{fangyixiang@cuhk.edu.cn}
  \end{tabular}
}

\begin{document}

\maketitle

\begin{abstract}
In emerging systems (e.g., social media and e-commerce platforms), data records are often drawn from heterogeneous sources, such as relational tables, text documents, image repositories, spatial databases, and knowledge graphs.
Accordingly, retrieving target records for question-answering (QA) tasks requires us to jointly exploit these heterogeneous sources.
However, most existing benchmarks are constructed from individual sources, and only a very few recent benchmarks have considered two or three sources.
To alleviate this issue, we introduce \ourDataset, a comprehensive benchmark with 857 QA pairs for record retrieval over five heterogeneous sources.
\ourDataset{} instantiates this setting with Yelp business records, each of which is grounded by multiple sources.
We build \ourDataset{} in an answer-driven manner: candidate records are first initialized with record-field constraints, then enriched through heterogeneous sources, and finally cross-verified across required sources before the natural-language question is retained.
We validate the benchmark through contradiction detection and human validation, and further evaluate sparse, dense, hybrid, late-interaction, and agentic retrievers under the same metrics.
The results show that \ourDataset{} is challenging: hybrid retrieval achieves the strongest Recall@10, Self-RAG achieves the best MRR@10, and all evaluated methods remain far from saturating the benchmark.
These findings indicate that \ourDataset{} provides an effective testbed for record retrieval over heterogeneous sources and leaves substantial room for future retrieval methods.
The benchmark dataset and source code are publicly available at \url{https://huggingface.co/datasets/hanchang02/HeterQA} and \url{https://github.com/hanchang02/HeterQA}, respectively.

\end{abstract}

\section{Introduction}
\label{sec:intro}

In emerging systems (e.g., social media and e-commerce platforms), natural-language interfaces have been increasingly employed to retrieve data records from heterogeneous sources, such as relational tables, textual documents, image repositories, spatial databases, and knowledge graphs (KGs).
For example, a user may ask for chicken wing restaurants that are highly rated, commented with positive words, close to the current location, and visually similar to a referenced venue.
To identify target records for such question-answering (QA) tasks, we have to jointly consider multiple heterogeneous sources, rather than individual sources which suffer from incomplete coverage and reliability.

Formally, in the record retrieval of heterogeneous sources, it often assumes there is a collection of records $\mathcal{R}$, and each record $r \in \mathcal{R}$ is drawn from multiple heterogeneous sources.
Given a natural language question $q$, it aims to return a ranked list of records matching with $q$.
A returned record is correct only when its source bundle satisfies all constraints expressed by $q$; retrieving an isolated text, image, spatial, or relational clue is insufficient if the record violates another required constraint.

\begin{table}[H]
\centering
\caption{Benchmark comparison where \textsc{P} denotes partial coverage.}
\label{tab:benchmark_compare}
\renewcommand{\arraystretch}{0.97}
\setlength{\tabcolsep}{3.0pt}
\footnotesize
\newcommand{\partialmark}{\textsc{P}}
\newcommand{\nomark}{$\times$}
\resizebox{0.85\textwidth}{!}{%
\begin{tabular}{@{}lcccccc@{}}
\toprule
\multirow{2}{*}{\textbf{Benchmark}} &
\multicolumn{5}{c}{\textbf{Data Sources}} &
\multicolumn{1}{c}{\textbf{Task Feature}} \\
\cmidrule(lr){2-6}\cmidrule(lr){7-7}
&
\makecell{\textbf{Relational}\\\textbf{Tables}} &
\makecell{\textbf{Text}\\\textbf{Documents}} &
\makecell{\textbf{Image}\\\textbf{Repositories}} &
\makecell{\textbf{Spatial}\\\textbf{Databases}} &
\makecell{\textbf{Knowledge}\\\textbf{Graphs}} &
\makecell{\textbf{Missing-Value}\\\textbf{Recovery}} \\
\midrule
HybridQA~\citep{hybridqa} & $\checkmark$ & $\checkmark$ & \nomark & \nomark & \nomark & \nomark \\
OTT-QA~\citep{ottqa} & $\checkmark$ & $\checkmark$ & \nomark & \nomark & \nomark & \nomark \\
MMQA~\citep{mmqa} & $\checkmark$ & $\checkmark$ & $\checkmark$ & \nomark & \nomark & \nomark \\
Spider2.0~\citep{spider2} & $\checkmark$ & \partialmark & \nomark & \nomark & \nomark & \nomark \\
BIRD~\citep{bird} & $\checkmark$ & \partialmark & \nomark & \nomark & \nomark & \nomark \\
STARK~\citep{stark} & $\checkmark$ & $\checkmark$ & \nomark & \nomark & $\checkmark$ & \nomark \\
\hline
\ourDataset{} & $\checkmark$ & $\checkmark$ & $\checkmark$ & $\checkmark$ & $\checkmark$ & $\checkmark$ \\
\bottomrule
\end{tabular}%
}
\end{table}

To evaluate record retrieval for QA tasks, several benchmarks have been developed.
HybridQA~\citep{hybridqa} aligns questions with Wikipedia tables and linked passages, and OTT-QA~\citep{ottqa} moves this table-text setting to open-domain QA over Wikipedia.
MMQA~\citep{mmqa} further combines text, tables, and images, but each question is built around a question-specific multimodal context rather than a full record collection for target-record retrieval. % M-BEIR ~\citep{uniir} standardizes universal multimodal retrieval over text, images, and interleaved text-image candidates; MM-Embed~\citep{mmembed} further studies retriever models under this M-BEIR setting.
Spider2.0~\citep{spider2} and BIRD~\citep{bird} stress realistic database reasoning and executable text-to-SQL tasks, yet their outputs are SQL programs or database answers rather than ranked target records.
STARK~\citep{stark} evaluates open retrieval over semi-structured knowledge bases with textual and relational information.
However, as summarized in Table~\ref{tab:benchmark_compare}, none of them has considered more than three sources, so they are still limited for benchmarking record retrieval for QA on multiple heterogeneous sources.
%
%none of these benchmarks evaluates open-domain record retrieval for QA over five sources: relational tables, text documents, image repositories, spatial databases, and knowledge graphs (KG).
%, but they have not yet evaluated the setting we target: open-domain record retrieval\footnote{\emph{Open-domain record retrieval} differs from open-domain QA: the method searches the full record collection $\mathcal{R}$ and returns records, rather than retrieving contexts for answer extraction.}, which aims to retrieve records jointly satisfying constraints from multiple heterogeneous sources.
%
Besides, they ignore the missing-value recovery: when answer sets start from a relational table, incomplete field values may leave valid target records outside the gold set.
For example, in a relational table, a record may have a missing value for the field like \texttt{parking}, but data from other sources like reviews and photos still indicate that parking is available.
Hence, it is desirable to develop a benchmark that is built from multiple heterogeneous sources with incomplete relational fields.

\begin{figure}[h]
    \centering
    \includegraphics[width=0.88\linewidth]{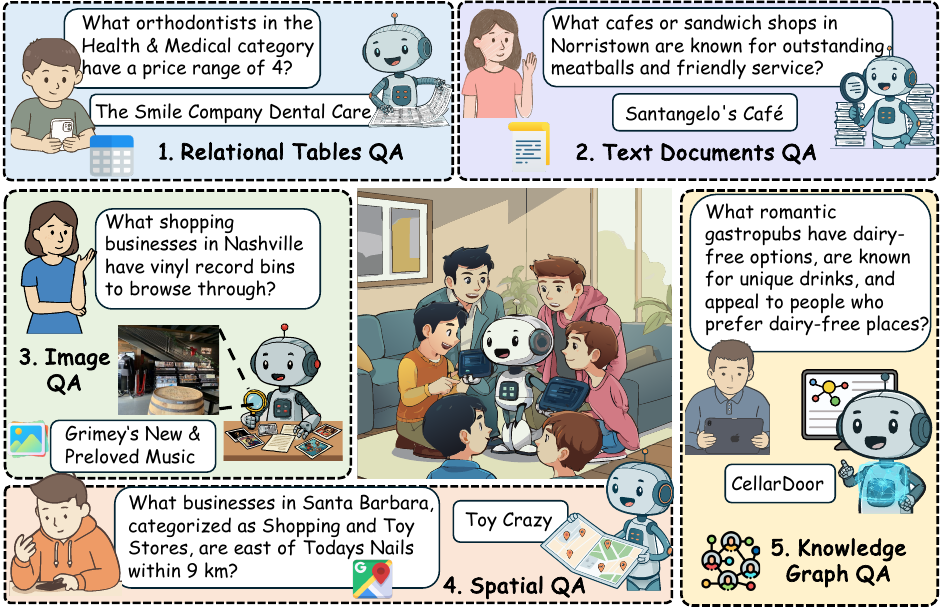}
    \caption{Representative questions from \ourDataset.}
    \label{fig:intro_examples}
\end{figure}

\textbf{Our work.}
We introduce \textbf{\ourDataset}, a benchmark with 857 QA pairs for record retrieval over five heterogeneous sources, and considering the missing-value recovery.
%including cases that require recovering valid target records from other sources.
%
We instantiate \ourDataset{} with the Yelp business record collection $\mathcal{R}$\footnote{\url{https://www.yelp.com/dataset} or \url{https://business.yelp.com/data/resources/open-dataset/}}.
%where each record contains 79 relational fields and 4 other source fields: text, image, spatial, and KG.
%
Particularly, each record contains 79 relational fields that are from the source of a relational table, and 4 other fields that are from the sources of text documents, image repositories, spatial databases, and KGs, respectively.
The benchmark is built in an answer-driven manner.
For each QA instance, we first sample a small set $F$ of relational-field constraints and execute them over $\mathcal{R}$ to obtain an initial answer set $A$.
Since relational-field values may be incomplete, we perform missing-value recovery: for each serialized field-value string $f\in F$, we search the other source fields of records outside $A$ and collect the recovered records as $C_f$.
This yields an enriched candidate set $\mathcal{C}=A\cup(\bigcup_{f\in F}C_f)$.
We then randomly select a record from $\mathcal{C}$, choose one of its other source fields, extract a source-specific constraint, and verbalize this constraint together with $F$ as a natural-language question $q$.
Finally, we verify which records in $\mathcal{C}$ satisfy all constraints in $q$; the candidate QA instance is retained only if the verified answer set is non-empty.

Further, we conduct human validation on sampled questions for naturalness, diversity, and practicality, obtaining non-negative rates of 97.0\%, 87.4\%, and 84.4\%, respectively.
Under the same evaluation metrics, we compare five kinds of retrieval methods: sparse retrieval, dense retrieval, hybrid retrieval (sparse + dense), late-interaction retrieval, and agentic retrieval.
In retrieval experiments, hybrid retrieval obtains the best Recall@10 at 32.78 and Self-RAG obtains the best MRR@10 at 25.26; together with human validation, these results show that \ourDataset{} is both high-quality and challenging for record retrieval over heterogeneous sources.

In summary, the key features of \ourDataset{} are:
\begin{itemize}[leftmargin=*, itemsep=1pt, topsep=2pt, parsep=0pt, partopsep=1pt]
    \item \textbf{Questions over five heterogeneous sources.}
    \ourDataset{} evaluates record retrieval over relational tables, text documents, image repositories, spatial databases, and KGs. Each question require methods to retrieve records that jointly satisfy all source-specific constraints.

    \item \textbf{Answer-driven construction with missing-value recovery.}
    \ourDataset{} first initializes candidate answer records from relational-field constraints, then recovers additional valid records from other source fields when relational-field values are incomplete. The final answer set is verified before the natural-language question is retained. Because each additional source can be introduced as a new source-specific constraint and verified through its support set, the workflow can be extended to more source fields.
    \item \textbf{Human-validated and challenging benchmark.}
    Human validation supports the naturalness, diversity, and practicality of the released questions, while retrieval experiments show that current methods remain far from saturating \ourDataset{}.
\end{itemize}

\section{Related Work}
\subsection{Benchmarks over Individual Sources}

Text-source benchmarks evaluate QA and retrieval over unstructured corpora.
Natural Questions~\citep{nq}, TriviaQA~\citep{triviaqa}, HotpotQA~\citep{hotpotqa}, MuSiQue~\citep{musique}, and MultiHop-RAG~\citep{multihoprag} cover open-domain, evidence-grounded, and multi-hop textual QA settings.
BEIR~\citep{beir} and KILT~\citep{kilt} make retrieval and provenance explicit over text corpora.
These benchmarks are valuable for passage retrieval and textual grounding, but they do not require a returned record to satisfy non-textual source constraints.

Text-to-SQL benchmarks evaluate natural-language querying over relational databases.
Spider~\citep{spider} is a representative cross-domain text-to-SQL benchmark, and BIRD~\citep{bird} extends database-grounded evaluation to larger database contents and external knowledge.
However, none of the above benchmarks has considered the retrieval from multiple heterogeneous sources.

\subsection{Benchmarks over Multiple Heterogeneous Sources}

Several QA and retrieval benchmarks jointly consider more than one source.
HybridQA~\citep{hybridqa} and OTT-QA~\citep{ottqa} combine relational tables with linked textual evidence.
MMQA~\citep{mmqa} combines text, tables, and images, but its questions are built around question-specific multimodal contexts rather than full-collection target-record retrieval.
M-BEIR~\citep{uniir} studies instruction-following multimodal retrieval over text, image, and interleaved text-image candidates, but it does not evaluate QA record retrieval over heterogeneous source bundles.
STARK~\citep{stark} evaluates open retrieval over semi-structured knowledge bases, where node entities are grounded in textual documents and relational graph structure.
Spider2.0~\citep{spider2} evaluates enterprise text-to-SQL workflows that require database metadata, SQL documentation, and project-level code.
BIRD-INTERACT~\citep{birdinteract} evaluates dynamic multi-turn text-to-SQL interactions with metadata, user simulation, and executable test cases.
However, as summarized in Table~\ref{tab:benchmark_compare}, none of them has considered more than three sources, so they are still limited for benchmarking record retrieval for QA on multiple heterogeneous sources.
Besides, they ignore the missing-value recovery as aforementioned.
Therefore, existing benchmarks are still insufficient for evaluating the record retrieval for QA over multiple hterogeneous sources.

%These benchmarks motivate our setting, but they do not jointly evaluate record retrieval over five heterogeneous sources with missing-value recovery.
%A separate line makes database benchmarks more realistic while retaining a database-centered target.
%
%Both differ from \ourDataset{} because they evaluate database workflows rather than retrieving target records whose constraints span relational and non-relational sources.
\section{Answer-driven Dataset Construction}
\suppressfloats[t]

\begin{figure}[h]
  \centering
  \includegraphics[width=0.94\textwidth]{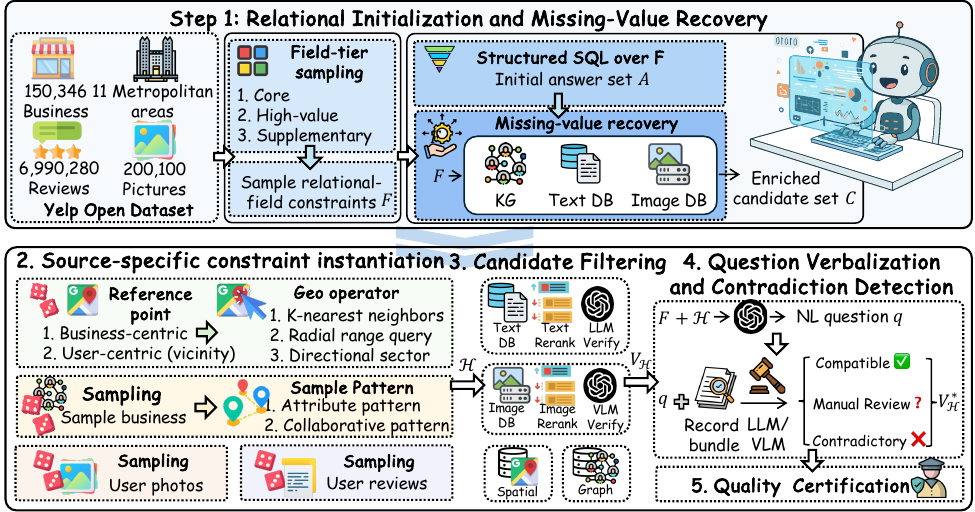}
  \setlength{\abovecaptionskip}{-2pt}
  \caption{Answer-driven dataset construction workflow.}
  \label{fig:main_process}
\end{figure}

Figure~\ref{fig:main_process} depicts our answer-driven construction workflow.
Each QA instance is constructed by a five-step approach:
We first initialize a set $F$ of relational-field constraints over $\mathcal{R}$, then expand the candidate set to $\mathcal{C}$, next instantiate source-specific constraints $\mathcal{H}$, and finally audit the verified answer set $\mathcal{V}_{\mathcal{H}}^\star$ for release.
In the following, we introduce these steps sequentially.

\subsection*{Step 1: Relational Initialization and Missing-Value Recovery}

For each QA instance, we sample a small set $F$ of relational-field constraints by first selecting one \textit{Core} field, then randomly selecting one or two fields from the union of the \textit{High-value} and \textit{Supplementary} tiers (Appendix~\ref{app:field_tiers}).
Each sampled field is paired with a value from $\mathcal{R}$, and the resulting key-value constraints are compiled into a structured SQL query over the relational table.
Executing this query over $\mathcal{R}$ gives the initial answer set $A$.

In the original relational table, since some fields' values are incomplete, we expand $A$ through missing-value recovery over three fields of non-relational sources, i.e., text, image, and KG.
In Yelp, text consists of reviews and tips, image consists of business photos, and KG denotes a feature-centric heterogeneous graph $\mathcal{G}$.\footnote{We construct $\mathcal{G}$ from Yelp reviews and tips by prompting an LLM to extract feature nodes and link them to users and businesses. Each link stores positive/negative sentiment to distinguish constraints such as ``good service'' from ``poor service,'' together with extraction confidence. The graph contains 237,082 nodes and 4,011,327 edges; details are in Appendix~\ref{app:kg_construction}.}
For each serialized field-value string $f\in F$, we search records outside $A$ through these source fields.
For text recovery, we search $f$ over all Yelp reviews and tips with text embeddings and rerank the matched texts; records (i.e., businesses in Yelp) above some threshold (see details in Appendix~\ref{app:construction_audit_ranking}) form $C_f^{\mathcal{T}}$.
Images are processed by a multimodal embedding model and a visual reranker, producing $C_f^{\mathcal{I}}$.
For KG recovery, we search $f$ over feature nodes in $\mathcal{G}$ and rerank the matched features; their one-hop neighboring business records form $C_f^{\mathcal{G}}$.
We set $C_f=C_f^{\mathcal{T}}\cup C_f^{\mathcal{I}}\cup C_f^{\mathcal{G}}$ and $\mathcal{C}=A\cup(\bigcup_{f\in F}C_f)$.
A recovered record is discarded if its relational fields explicitly contradict any constraint in $F$.

\subsection*{Step 2: Source-Specific Constraint Instantiation}

After forming $\mathcal{C}$, we sample an anchor record and select one or two non-relational source fields, each yielding a source-specific constraint $h$.
For text, $h$ is generated by sampling a review or tip and rewriting it into a concise requirement.
For image, $h$ verbalizes a visible property from a sampled business photo, such as sea view.
For spatial information, we use the anchor record to define a local search area, then choose either a nearby business or a simulated user location sampled from that area as the reference point.
We pair this point with a spatial operator to form $h$, such as radius (e.g., within 2 km of a cafe), direction (e.g., north of a museum), or nearest-neighbor selection (e.g., the closest laundromat to the user location).
For KG, we inspect the graph neighborhoods of candidate records in $\mathcal{C}$ and sample two types of feature patterns as $h$.
Attribute patterns use direct business--feature links, such as requiring ``friendly staff''.
Collaborative patterns use shared user connections, such as venues with ``quiet ambience'' that are also favored by users who praise ``desserts''.

\subsection*{Step 3: Candidate Filtering}

Recall that in Step 2, the constraint $h$ is instantiated from one specific anchor record, but it may be satisfied by many other candidate records in $\mathcal C$, so we have to exclude those records that do not satisfy $h$ from $\mathcal C$. 
%
%Although $h$ is instantiated from one anchor record, candidate filtering applies $h$ to all records in $\mathcal{C}$ and keeps only the records that support it.
%
Specifically, for each selected $h$, we define a support set $\mathcal{V}_h\subseteq\mathcal{C}$.
For non-spatial $h$, verification is cross-source because $h$ may be supported outside the source field that generated it.
Text- and image-derived $h$ are checked against the other source field through embedding retrieval, reranking, and LLM/VLM judgment using the relevance prompts in Figures~\ref{fig:vlm_judge_prompt} and ~\ref{fig:relevance_prompt}; KG-derived $h$ is checked on the graph and against text/image source fields with the same retrieval-and-judgment pipeline.
A record enters $\mathcal{V}_h$ if any applicable check passes; for spatial $h$, $\mathcal{V}_h$ is obtained deterministically from the reference point and geo operator.
For the selected constraint set $\mathcal{H}$, we keep records in $\mathcal{V}_{\mathcal{H}}=\bigcap_{h\in\mathcal{H}}\mathcal{V}_h$ and carry their verification confidences forward for ranking.

\subsection*{Step 4: Question Generation and Contradiction Detection}

After obtaining $\mathcal{V}_{\mathcal{H}}$, we invoke an LLM to verbalize the relational constraints $F$ and source-specific constraints $\mathcal{H}$ into a natural-language question $q$.
However, some records in $\mathcal{V}_{\mathcal{H}}$ may still contain contradictory evidence elsewhere in their source bundles, so we perform contradiction detection against the generated question.
For each record in $\mathcal{V}_{\mathcal{H}}$, the detector retrieves query-relevant reviews/tips and photos, then estimates text and image contradiction ratios with the prompts in Figures~\ref{fig:text_contradiction_prompt} and~\ref{fig:visual_contradiction_prompt}.
Let $\mathcal{V}_{\mathcal{H}}^\star\subseteq\mathcal{V}_{\mathcal{H}}$ be the records that pass contradiction detection: records above the contradiction threshold are removed, and unresolved cases are sent to manual review.
The final set $\mathcal{V}_{\mathcal{H}}^\star$ is the verified answer set and is ranked by the composite score in Appendix~\ref{app:construction_audit_ranking}.

\subsection*{Step 5: Human Validation}

Human validation is used in two stages.
First, unresolved cases from contradiction detection are manually checked before finalizing the verified answer set $\mathcal{V}_{\mathcal{H}}^\star$.
Second, to assess question quality, we sample 200 questions with subset-proportional stratified sampling and ask 13 graduate annotators, including PhD and MPhil students, to rate question naturalness, diversity, and practicality.
Following STARK~\citep{stark}, we also compute word entropy and Type-Token Ratio (TTR) to measure query diversity.
Table~\ref{tab:dataset_stats_master} reports dataset composition, query-diversity metrics, and human-validation results.
The resulting dataset contains 857 QA pairs across ten source-composition subsets.
The released questions obtain a word entropy of 8.987 and a TTR of 0.160, and the non-negative rates for naturalness, diversity, and practicality are 97.0\%, 87.4\%, and 84.4\%, respectively.
These results indicate that the released questions are generally understandable, varied, and practically meaningful.
Metric definitions are in Appendix~\ref{app:quality_validation}, and construction settings are in Appendix~\ref{app:construction_settings}.
\begin{table}[t!]
\centering

\caption{Dataset statistics and quality certification for \ourDataset, where $\mathcal{T}$, $\mathcal{I}$, $\mathcal{P}$, and $\mathcal{G}$ denote review text, photos, spatial information, and KG, respectively.}
\label{tab:dataset_stats_master}
\label{tab:quality_audit}
\renewcommand{\arraystretch}{1.05}
\setlength{\tabcolsep}{2.5pt}
\scriptsize
\resizebox{\textwidth}{!}{%
\begin{tabular}{@{}l|cccc|cccccc|c@{}}
\toprule
\multirow{4}{*}{\textbf{Metric}} & \multicolumn{4}{c|}{\textbf{Relational + One Source Field}} & \multicolumn{6}{c|}{\textbf{Relational + Two Source Fields}} & \multirow{4}{*}{\textbf{Overall}} \\
 & \multicolumn{4}{c|}{\small \textit{Composition:} $S+\{h_1\}$} & \multicolumn{6}{c|}{\small \textit{Composition:} $S+\{h_1,h_2\}$} & \\
\cmidrule(lr){2-5} \cmidrule(lr){6-11}
 & \textbf{$Q_{S+\mathcal{T}}$} & \textbf{$Q_{S+\mathcal{I}}$} & \textbf{$Q_{S+\mathcal{P}}$} & \textbf{$Q_{S+\mathcal{G}}$} & \textbf{$Q_{S+\mathcal{T,I}}$} & \textbf{$Q_{S+\mathcal{T,P}}$} & \textbf{$Q_{S+\mathcal{T,G}}$} & \textbf{$Q_{S+\mathcal{I,P}}$} & \textbf{$Q_{S+\mathcal{I,G}}$} & \textbf{$Q_{S+\mathcal{P,G}}$} & \\
\midrule
\textbf{\# Questions} & 96 & 66 & 138 & 46 & 74 & 107 & 72 & 133 & 44 & 81 & 857 \\
\textbf{Avg. Length} & 21.3 & 12.8 & 16.9 & 20.7 & 22.9 & 28.5 & 32.3 & 18.5 & 24.1 & 27.4 & 22.2 \\
\textbf{Avg. Verified Answers} & 2.8 & 2.7 & 2.4 & 3.1 & 2.6 & 1.5 & 2.9 & 2.2 & 1.4 & 1.5 & 2.3 \\
\bottomrule
\end{tabular}
}
\vspace{0.5em}

\begin{minipage}[t]{0.34\textwidth}
\centering
\small
\begin{tabular}{@{}ll@{}}
\toprule
\multicolumn{2}{c}{\textbf{Query Diversity}} \\
\midrule
\textbf{Metric} & \textbf{Overall} \\
\midrule
Word Entropy~\cite{stark} & 8.987 \\
Type-Token Ratio (TTR)~\cite{stark} & 0.160 \\
\bottomrule
\end{tabular}
\end{minipage}\hfill
\begin{minipage}[t]{0.36\textwidth}
\centering
\small
\begin{tabular}{@{}lccc@{}}
\toprule
\multicolumn{4}{c}{\textbf{Human Quality}} \\
\midrule
\textbf{Dimension} & \textbf{Positive} & \textbf{Non-neg.} & \textbf{Mean} \\
\midrule
Naturalness & 87.5\% & 97.0\% & 4.17 \\
Diversity & 64.8\% & 87.4\% & 3.62 \\
Practicality & 67.4\% & 84.4\% & 3.82 \\
\bottomrule
\end{tabular}
\end{minipage}
\vspace{-0.1em}
\end{table}

\section{Experiments}
\label{sec:experiments}

\subsection{Experimental Setup}

We evaluate all methods on the 857 QA pairs in \ourDataset{}; each pair provides a question $q$ and a verified answer set $\mathcal{V}_{\mathcal{H}}^\star$.
Given $q$, a retriever searches $\mathcal{R}$ using serialized relational fields, Yelp reviews/tips, and business photos, then returns a ranked record list.
Location metadata is included in the serialized relational fields, so retrievers can access location cues without a separate spatial index.

We evaluate five method families. 
(a) Sparse retrieval uses BM25~\citep{bm25}; it retrieves from the three sources and fuses source-specific record rankings with RRF~\citep{rrf}.
(b) Dense retrieval uses DPR~\citep{dpr}, ANCE~\citep{ance}, KALM~\citep{kalmv2}, Llama-Embed-Nemotron-8B~\citep{llamaembednemotron}, Qwen3-Embedding-8B~\citep{qwen3embedding}, and Qwen3-VL-Embedding-8B~\citep{qwen3vlembedding}, following the same fusion procedure.
(c) Hybrid retrieval adds BM25 as a parallel branch to each dense retriever and fuses sparse and dense record rankings with another RRF step.
(d) Late-interaction retrieval includes ColBERT~\citep{colbert} and ColPali~\citep{colpali}; ColPali retrieves rendered Yelp-record pages and maps page hits back to records.
(e) Agentic retrieval includes ReAct~\citep{react} and Self-RAG~\citep{selfrag}. 
For photos, caption-based retrieval embeds VLM-generated captions, while direct-visual retrieval embeds the original business photos with Qwen3-VL-Embedding-8B.
We report results without reranking (w/o R) and with reranking (w/ R), where w/ R uses Qwen3-VL-Reranker-8B~\citep{qwen3vlembedding}.
Hit@5 measures whether the top-5 list contains at least one record in $\mathcal{V}_{\mathcal{H}}^\star$, Recall@10 measures coverage of $\mathcal{V}_{\mathcal{H}}^\star$, MRR@10 measures the rank of the first correct record, and latency is average end-to-end question time.
All experiments are conducted on a Linux server with Intel Xeon 2.0 GHz CPUs, 1024 GB RAM, and eight NVIDIA GeForce RTX A5000 GPUs.
The two agentic workflows use Qwen3.5-9B~\citep{qwen35} as the backend model, with a 100K-token context budget and 4 concurrent workers. 
Vector-based retrieval is served by OceanBase HNSW indexes over three source stores: about 7.89M text entries, 0.20M photo entries, and 0.15M serialized relational-field entries.
Detailed retrieval settings and agentic procedures are provided in Appendix~\ref{app:retrieval_algorithm} and Appendix~\ref{app:agentic_methods}.

\newcommand{\metricgain}[1]{{\textcolor{blue!60!black}{\scriptsize\,#1}}}
\providecommand{\hybridbest}[1]{\underline{#1}}
\providecommand{\hybridsecond}[1]{\underline{\underline{#1}}}

\begin{table*}[htbp]
\centering
\caption{Overall retrieval results on \ourDataset{}. \textbf{Bold black values} mark the best raw score in each column. \textcolor{blue!60!black}{Blue offsets} show gains from hybrid retrieval over the corresponding base retriever, with \textcolor{blue!60!black}{\textbf{bold blue offsets}} marking the largest gain. \hybridbest{Single} and \hybridsecond{double} underlines mark the best and second-best hybrid retrieval results.}
\label{tab:main_result}
\renewcommand{\arraystretch}{1.12}
\setlength{\tabcolsep}{4.0pt}
\footnotesize
\resizebox{\textwidth}{!}{%
\begin{tabular}{@{}lcccccccc@{}}
\toprule
\multirow{2}{*}{Method} & \multicolumn{4}{c}{w/o R} & \multicolumn{4}{c}{w/ R} \\
\cmidrule(lr){2-5} \cmidrule(lr){6-9}
& Hit@5 $\uparrow$ & Recall@10 $\uparrow$ & MRR@10 $\uparrow$ & Lat. $\downarrow$ & Hit@5 $\uparrow$ & Recall@10 $\uparrow$ & MRR@10 $\uparrow$ & Lat. $\downarrow$ \\
\midrule
BM25 & 18.09 & 17.89 & 11.12 & 0.93 & 28.00 & \textbf{27.21} & 16.80 & 1.50 \\
DPR & 1.40 \metricgain{\textbf{+9.57}} & 1.48 \metricgain{\textbf{+11.20}} & 0.79 \metricgain{+5.07} & 0.08 \metricgain{+1.07s} & 3.15 \metricgain{\textbf{+13.54}} & 2.01 \metricgain{\textbf{+15.96}} & 1.35 \metricgain{+8.23} & \textbf{0.89} \metricgain{+4.70s} \\
ANCE & 6.77 \metricgain{+8.63} & 6.15 \metricgain{+8.93} & 4.58 \metricgain{+5.70} & \textbf{0.07} \metricgain{+1.09s} & 10.85 \metricgain{+11.09} & 7.48 \metricgain{+14.59} & 5.82 \metricgain{+8.76} & 0.92 \metricgain{+4.83s} \\
KALM & 18.09 \metricgain{+4.90} & 15.96 \metricgain{+6.06} & 10.02 \metricgain{+5.22} & 0.41 \metricgain{+0.83s} & 24.04 \metricgain{+5.48} & 20.86 \metricgain{+7.79} & 12.06 \metricgain{+8.66} & 0.99 \metricgain{+4.61s} \\
Llama & \hybridbest{21.94} \metricgain{\hybridbest{+3.62}} & \hybridbest{21.44} \metricgain{\hybridbest{+3.55}} & \hybridbest{11.20} \metricgain{\hybridbest{\textbf{+7.21}}} & 0.25 \metricgain{+0.98s} & \hybridbest{26.25} \metricgain{\hybridbest{+6.65}} & \hybridbest{24.28} \metricgain{\hybridbest{+8.50}} & \hybridbest{12.83} \metricgain{\hybridbest{\textbf{+10.71}}} & 0.98 \metricgain{+4.58s} \\
Qwen3 & 21.59 \metricgain{+2.10} & \hybridsecond{19.61} \metricgain{\hybridsecond{+4.39}} & \hybridsecond{11.81} \metricgain{\hybridsecond{+5.95}} & 0.44 \metricgain{+0.86s} & \hybridsecond{24.27} \metricgain{\hybridsecond{+5.60}} & \hybridsecond{21.64} \metricgain{\hybridsecond{+9.45}} & 13.00 \metricgain{+8.69} & 0.95 \metricgain{+4.58s} \\
Qwen3-VL (caption)$^\dagger$ & \hybridsecond{18.90} \metricgain{\hybridsecond{+5.48}} & 17.78 \metricgain{+5.03} & 12.05 \metricgain{+5.21} & 0.08 \metricgain{+1.04s} & \hybridsecond{25.67} \metricgain{\hybridsecond{+4.20}} & 21.71 \metricgain{+7.42} & 14.37 \metricgain{+8.37} & 1.02 \metricgain{+4.50s} \\
Qwen3-VL (photo)$^\ddagger$ & 18.55 \metricgain{+5.37} & 17.32 \metricgain{+5.27} & 11.69 \metricgain{+5.29} & 0.34 \metricgain{+0.79s} & 24.62 \metricgain{+5.13} & 20.89 \metricgain{+8.91} & \hybridsecond{13.69} \metricgain{\hybridsecond{+9.05}} & 1.00 \metricgain{+4.56s} \\
\midrule
ColBERT & 20.54 & 19.92 & 13.84 & 0.45 & 26.84 & 27.09 & 17.02 & 2.08 \\
ColPali & 0.82 & 0.95 & 0.41 & 1.16 & \multicolumn{4}{c}{--} \\
\midrule
ReAct & 26.72 & 23.42 & 21.61 & 200.93 & 28.70 & 24.06 & 22.88 & 132.07 \\
Self-RAG & \textbf{29.17} & \textbf{25.33} & \textbf{23.97} & 170.93 & \textbf{30.69} & 25.77 & \textbf{25.26} & 108.48 \\
\bottomrule
\end{tabular}%
}
\vspace{1pt}
\begin{minipage}{0.98\textwidth}
\scriptsize
\textit{Notes.}
$^\dagger$ The caption setting embeds VLM-generated captions of business photos.
$^\ddagger$ The photo setting embeds original business photos directly with Qwen3-VL-Embedding-8B. 
Latency offsets denote added seconds.

\end{minipage}
\end{table*}

\begin{figure}[t!]
\centering
\input{sections/figures/generated/subset_atlas_rows}

\definecolor{subsetgainblue}{RGB}{73,110,177}
\definecolor{subsetgainred}{RGB}{186,92,92}
\definecolor{subsetna}{RGB}{232,234,238}

\newcommand{\subsetheatcell}[1]{}
\newcommand{\subsetnacell}{\cellcolor{subsetna}\makebox[2.85em][c]{\textcolor{gray}{\scriptsize N/A}}}
\newcommand{\subsetlegendbox}[1]{\textcolor{#1}{\rule{8pt}{8pt}}}
\newcommand{\subsetqheader}[1]{$\scriptstyle Q_{#1}$}
\newcommand{\subsetgaincell}[2]{%
  \begingroup
  \pgfmathsetmacro{\subsetvalue}{#1}
  \pgfmathsetmacro{\subsetcap}{#2}
  \pgfmathsetmacro{\subsetratio}{min(abs(\subsetvalue)/\subsetcap, 1)}
  \pgfmathsetmacro{\subsetmix}{15 + 75*\subsetratio}
  \ifdim \subsetvalue pt > 0pt
    \edef\subsetcolorcmd{\noexpand\cellcolor{subsetgainblue!\subsetmix!white}}%
  \else\ifdim \subsetvalue pt < 0pt
    \edef\subsetcolorcmd{\noexpand\cellcolor{subsetgainred!\subsetmix!white}}%
  \else
    \edef\subsetcolorcmd{\noexpand\cellcolor{white}}%
  \fi\fi
  \subsetcolorcmd\makebox[2.85em][c]{\scriptsize\pgfmathprintnumber[fixed,precision=1,showpos]{\subsetvalue}}%
  \endgroup
}
\newcommand{\subsetpaneltitle}[1]{\vspace{0.05em}{\scriptsize\textbf{#1}}\par\vspace{0.15em}}
\newcommand{\subsetlegend}[1]{%
  {\tiny
  \subsetlegendbox{subsetgainred!85!white}\,loss
  \quad
  \subsetlegendbox{white}\,0
  \quad
  \subsetlegendbox{subsetgainblue!85!white}\,gain
  \quad
  \subsetlegendbox{subsetna}\,N/A
  \quad
  darker = larger change, clipped at $\pm #1$ points}}

{\scriptsize
\setlength{\tabcolsep}{1.35pt}
\renewcommand{\arraystretch}{0.95}
\renewcommand{\subsetheatcell}[1]{\subsetgaincell{#1}{15}}
\subsetpaneltitle{(a) Reranking gain: $\mathrm{Recall@10}(\mathrm{w/ R}) - \mathrm{Recall@10}(\mathrm{w/o R})$}
\resizebox{\linewidth}{!}{%
\begin{tabular}{@{}l*{10}{c}@{}}
\toprule
Method
& \subsetqheader{S+\mathcal{T}}
& \subsetqheader{S+\mathcal{I}}
& \subsetqheader{S+\mathcal{P}}
& \subsetqheader{S+\mathcal{G}}
& \subsetqheader{S+\mathcal{T,I}}
& \subsetqheader{S+\mathcal{T,P}}
& \subsetqheader{S+\mathcal{T,G}}
& \subsetqheader{S+\mathcal{I,P}}
& \subsetqheader{S+\mathcal{I,G}}
& \subsetqheader{S+\mathcal{P,G}} \\
\midrule
\multicolumn{11}{@{}l}{\scriptsize\textbf{Sparse / Dense}} \\
\SubsetWRGainRowsSparseDense
\midrule
\multicolumn{11}{@{}l}{\scriptsize\textbf{Hybrid Retrieval}} \\
\SubsetWRGainRowsHybrid
\midrule
\multicolumn{11}{@{}l}{\scriptsize\textbf{Late-Interaction / Agentic}} \\
\SubsetWRGainRowsLateAgentic
\bottomrule
\end{tabular}}
\vspace{0.12em}
\subsetlegend{15}
}

\vspace{0.45em}

{\scriptsize
\setlength{\tabcolsep}{1.35pt}
\renewcommand{\arraystretch}{0.95}
\renewcommand{\subsetheatcell}[1]{\subsetgaincell{#1}{31}}
\subsetpaneltitle{(b) Hybrid retrieval gain in Recall@10}
\resizebox{\linewidth}{!}{%
\begin{tabular}{@{}l*{10}{c}@{}}
\toprule
Dense retriever
& \subsetqheader{S+\mathcal{T}}
& \subsetqheader{S+\mathcal{I}}
& \subsetqheader{S+\mathcal{P}}
& \subsetqheader{S+\mathcal{G}}
& \subsetqheader{S+\mathcal{T,I}}
& \subsetqheader{S+\mathcal{T,P}}
& \subsetqheader{S+\mathcal{T,G}}
& \subsetqheader{S+\mathcal{I,P}}
& \subsetqheader{S+\mathcal{I,G}}
& \subsetqheader{S+\mathcal{P,G}} \\
\midrule
\multicolumn{11}{@{}l}{\scriptsize\textbf{w/o R}} \\
\SubsetHybridRowsWoR
\midrule
\multicolumn{11}{@{}l}{\scriptsize\textbf{w/ R}} \\
\SubsetHybridRowsWR
\bottomrule
\end{tabular}}
\vspace{0.12em}
\subsetlegend{31}
}

\caption{Source-combination heatmaps for Recall@10. Panel (a) reports the gain from reranking; panel (b) reports the gain from hybrid retrieval over the corresponding dense retriever. Blue indicates positive gain, red indicates loss, and gray denotes N/A.}
\label{fig:subset_atlas}
\end{figure}

\subsection{Overall Retrieval Results}

Table~\ref{tab:main_result} evaluates whether methods can retrieve the verified answer set $\mathcal{V}_{\mathcal{H}}^\star$ from the full record collection $\mathcal{R}$.
Given the natural-language question $q$, each method returns a ranked list of records, and all metrics compare this list with $\mathcal{V}_{\mathcal{H}}^\star$.
The results reveal four main patterns.

\noindent\textbf{Observation 1. Current methods recover only a limited part of the verified answer set.}
The best Recall@10 is 32.78, achieved by hybrid retrieval with Llama and reranking.
This shows that even the strongest setting retrieves only part of $\mathcal{V}_{\mathcal{H}}^\star$.
Self-RAG obtains the best raw MRR@10 at 25.26, but its Recall@10 remains below the best hybrid retrieval result.
Thus, current methods still struggle both to retrieve enough records in $\mathcal{V}_{\mathcal{H}}^\star$ and to rank verified records near the top.

\noindent\textbf{Observation 2. Hybrid retrieval contributes larger gains than reranking.}
Hybrid retrieval consistently improves over the corresponding base retrievers.
For dense retrievers, the average Recall@10 gain from reranking is 2.73 points, while the average Recall@10 gain from hybrid retrieval under w/ R is 10.37 points.
The same pattern holds for MRR@10: reranking adds 1.57 points on average, while hybrid retrieval adds 8.92 points.
For Llama, reranking improves Recall@10 from 21.44 to 24.28, whereas hybrid retrieval further increases it to 32.78.
These results suggest that expanding the retrieved record set through hybrid retrieval matters more than reranking the initial dense-retrieval results.

\begin{figure*}[t]
\centering
\definecolor{agentblue}{RGB}{63,105,170}
\definecolor{agentorange}{RGB}{199,112,54}
\begin{tikzpicture}
\begin{groupplot}[
  group style={group size=3 by 1, horizontal sep=0.82cm},
  width=0.31\textwidth, height=4.1cm,
  grid=both, grid style={draw=gray!15},
  tick label style={font=\scriptsize},
  label style={font=\scriptsize},
  title style={font=\scriptsize\bfseries},
  enlarge x limits=0.08,
]
\nextgroupplot[title={(a) Hit@5 vs Turn}, xlabel={Turn}, ylabel={Score (\%)}, ymin=25.5, ymax=30.2, xtick={5,10,15,20}]
\addplot+[agentblue, mark=*, thick] coordinates {(5.00,26.14) (10.00,26.72) (15.00,27.54) (20.00,27.77)};
\addplot+[agentorange, mark=*, thick] coordinates {(5.00,28.24) (10.00,29.17) (15.00,29.64) (20.00,28.94)};
\nextgroupplot[title={(b) Recall@10 vs Turn}, xlabel={Turn}, ymin=22, ymax=26.2, xtick={5,10,15,20}]
\addplot+[agentblue, mark=*, thick] coordinates {(5.00,22.73) (10.00,23.42) (15.00,23.97) (20.00,23.52)};
\addplot+[agentorange, mark=*, thick] coordinates {(5.00,23.57) (10.00,25.33) (15.00,25.30) (20.00,24.63)};
\nextgroupplot[title={(c) MRR@10 vs Turn}, xlabel={Turn}, ymin=20, ymax=25.6, xtick={5,10,15,20}]
\addplot+[agentblue, mark=*, thick] coordinates {(5.00,20.90) (10.00,21.61) (15.00,22.51) (20.00,22.07)};
\addplot+[agentorange, mark=*, thick] coordinates {(5.00,21.91) (10.00,23.97) (15.00,24.63) (20.00,23.05)};
\end{groupplot}
\node[anchor=north, yshift=-0.78cm, font=\tiny] at (group c2r1.south)
{\tikz{\draw[agentblue, thick] (0,0)--(0.34,0); \fill[agentblue] (0.17,0) circle (1.25pt);} ReAct
\hspace{1.0em}
\tikz{\draw[agentorange, thick] (0,0)--(0.34,0); \fill[agentorange] (0.17,0) circle (1.25pt);} Self-RAG};
\end{tikzpicture}
\caption{Agentic retrieval quality under w/o R as the maximum number of turns changes. Panels follow the metric order in Table~\ref{tab:main_result}. Self-RAG peaks around 10--15 turns, indicating that additional turns do not reliably recover more verified records.}
\label{fig:agent_budget}
\end{figure*}
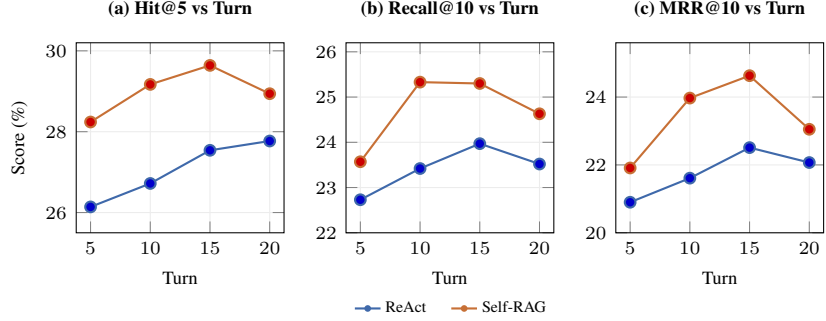

\noindent\textbf{Observation 3. Agentic retrieval improves top-rank metrics but is not a recall solution.}
Self-RAG obtains the best raw Hit@5 and MRR@10, reaching 30.69 and 25.26 w/ R, respectively.
However, its Recall@10 is 25.77, still below the best hybrid retrieval result of 32.78.
The gain also comes with a large latency cost: Self-RAG takes 108.48 seconds per question w/ R, compared with 5.56 seconds for hybrid retrieval with Llama under the same setting.
Figure~\ref{fig:agent_budget} further shows that increasing the turn budget does not monotonically improve retrieval.
Thus, agentic retrieval can improve early correct hits, but higher latency and more turns do not reliably recover more records in $\mathcal{V}_{\mathcal{H}}^\star$.

\noindent\textbf{Observation 4. Late-interaction retrieval does not overcome the record-level difficulty.}
ColBERT reaches 27.09 Recall@10 w/ R, which is close to BM25 at 27.21 but does not improve over it.
This suggests that token-level interaction over text-like content is not enough to solve \ourDataset{}, even when serialized fields, reviews/tips, and photo captions are available.
ColPali performs much worse: after each Yelp business is rendered into PDF pages, it reaches only 0.95 Recall@10 w/o R.
These results show that current late-interaction baselines do not directly address the target-record retrieval setting, where a method must return records satisfying constraints across heterogeneous sources.

\noindent\textbf{Observation 5. Visual photo embeddings remove the need for caption conversion.}
Qwen3 reaches 21.64 Recall@10 w/ R on caption-based photo retrieval, comparable to Qwen3-VL (caption) at 21.71 and Qwen3-VL (photo) at 20.89.
After hybrid retrieval, the Recall@10 scores remain close: 31.09 for Qwen3, 29.13 for Qwen3-VL (caption), and 29.80 for Qwen3-VL (photo).
This suggests that visual photo embeddings achieve comparable retrieval quality to caption-based embeddings while avoiding the extra VLM captioning step.

\subsection{Source-Combination Analysis}

Figure~\ref{fig:subset_atlas} shows how retrieval behavior changes across source-combination subsets.
Panel (a) reports reranking gains, and Panel (b) reports hybrid retrieval gains over the corresponding dense retriever.

\noindent\textbf{Observation 6. Reranking gains vary across source combinations and methods.}
Reranking is not uniformly beneficial across subsets.
For BM25, reranking gives large Recall@10 gains on text-bearing combinations, including +14.78 on $Q_{S+\mathcal{T,I}}$, +12.62 on $Q_{S+\mathcal{T,P}}$, and +12.22 on $Q_{S+\mathcal{T,G}}$, but only +3.09 on $Q_{S+\mathcal{P,G}}$.
A similar dependence appears for ColBERT, which gains +11.59 on $Q_{S+\mathcal{T,I}}$ and +10.23 on $Q_{S+\mathcal{I,P}}$, but loses 1.52 points on $Q_{S+\mathcal{G}}$.
These patterns show that reranking helps when the retrieved candidates already contain compatible source cues, but it is not a uniform fix for source-combination difficulty.

\noindent\textbf{Observation 7. Hybrid retrieval gives broad gains, but spatial-KG cases remain hard.}
Hybrid retrieval improves Recall@10 across several dense retrievers, especially on subsets involving text documents.
Under w/ R, adding BM25 improves DPR by +30.00 on $Q_{S+\mathcal{T,P}}$, +27.62 on $Q_{S+\mathcal{T,I}}$, and +23.47 on $Q_{S+\mathcal{T,G}}$.
ANCE shows the same trend, with +27.10 on $Q_{S+\mathcal{T,P}}$ and +18.57 on $Q_{S+\mathcal{T,G}}$.
For stronger retrievers, the gains remain visible: Qwen3 gains +13.79 on $Q_{S+\mathcal{T,P}}$, and Qwen3-VL (photo) gains +11.75 on $Q_{S+\mathcal{T,G}}$.
However, gains on $Q_{S+\mathcal{P,G}}$ are smaller for most retrievers, such as +4.32 for DPR, +4.94 for ANCE, +3.26 for Llama, and +0.79 for Qwen3-VL (photo).
Thus, hybrid retrieval helps recover records missed by dense retrievers, but questions combining spatial constraints and KGs remain difficult.

% \input{sections/figures/failure_taxonomy}
% \input{sections/tables/evidence_grounding_cases}

% \subsection{Failure Analysis}

% \noindent\textbf{Observation 7. Most errors come from missing source support or failed constraint composition.}
% %
% Figure~\ref{fig:failure_taxonomy} summarizes 50 manually reviewed retrieval failures.
% %
% Missing source support accounts for 26 cases, and failed composition of source-specific constraints accounts for another 12 cases.
% %
% Moreover, 29 of the 50 failures are Recall@10 misses, meaning that records in $\mathcal{V}_{\mathcal{H}}^\star$ are absent from the returned top-10 list rather than merely ranked too low.
% %
% Table~\ref{tab:evidence_grounding_cases} shows representative cases where the missed constraint comes from reviews/tips, photos, spatial constraints, or KGs rather than a single relational field.

\subsection{RAG Experiments}

We evaluate downstream RAG record selection with four generator backends: Qwen3.5-9B, Qwen3.5-27B~\citep{qwen35}, Gemma4-31B~\citep{gemma4}, and GLM-4.6V-Flash~\citep{glmv}.
For each question, the retriever first produces a top-30 record list after record-level collapse; the generator LLM receives these records with associated source content and outputs the selected records.
Detailed RAG evaluation setup and prompt templates are provided in Appendix~\ref{app:rag_details} and Appendix~\ref{app:examples}.

\providecommand{\metricgainpos}[1]{{\textcolor{blue!60!black}{\scriptsize\,#1}}}
\providecommand{\metricloss}[1]{{\textcolor{red!65!black}{\scriptsize\,#1}}}
\providecommand{\hybridbest}[1]{\underline{#1}}
\providecommand{\hybridsecond}[1]{\underline{\underline{#1}}}
\begin{table}[t!]
\centering
\caption{RAG record-selection results on \ourDataset{}. Values are business-id Macro-F1 (\%). \textbf{Bold black values} mark the best raw score in each column. \textcolor{blue!60!black}{Blue offsets} and \textcolor{red!65!black}{red offsets} show gains and regressions from hybrid retrieval over the corresponding base retriever, with \textcolor{blue!60!black}{\textbf{bold blue offsets}} marking the largest gain. \hybridbest{Single} and \hybridsecond{double} underlines mark the best and second-best hybrid retrieval results. Generation errors are counted in the 857-QA-pair denominator; ColPali has only w/o R runs.}
\label{tab:rag_result}
\renewcommand{\arraystretch}{1.08}
\setlength{\tabcolsep}{4.2pt}
\footnotesize
\resizebox{\textwidth}{!}{%
\begin{tabular}{@{}lcccccccc@{}}
\toprule
\multirow{2}{*}{Retriever} & \multicolumn{2}{c}{Qwen3.5-9B} & \multicolumn{2}{c}{Qwen3.5-27B} & \multicolumn{2}{c}{Gemma4-31B} & \multicolumn{2}{c}{GLM-4.6V-Flash} \\
\cmidrule(lr){2-3} \cmidrule(lr){4-5} \cmidrule(lr){6-7} \cmidrule(lr){8-9}
 & w/o R & w/ R & w/o R & w/ R & w/o R & w/ R & w/o R & w/ R \\
\midrule
BM25 & 7.88 & 8.48 & 8.25 & 9.28 & \textbf{12.61} & \textbf{13.36} & 8.27 & 8.00 \\
DPR & 1.01 \metricgainpos{+0.17} & 1.02 \metricgainpos{+0.29} & 1.31 \metricgainpos{+0.02} & 1.38 \metricgainpos{+0.34} & 1.24 \metricgainpos{+0.21} & 1.21 \metricgainpos{+0.79} & 1.62 \metricloss{-0.08} & 1.49 \metricgainpos{+0.08} \\
ANCE & 3.59 \metricgainpos{+0.64} & 3.20 \metricgainpos{\textbf{+1.91}} & 4.22 \metricgainpos{+0.10} & 4.24 \metricgainpos{+1.33} & 3.99 \metricgainpos{+0.98} & 4.45 \metricgainpos{+2.15} & 4.42 \metricgainpos{+0.53} & 4.56 \metricgainpos{\textbf{+1.18}} \\
KALM & 7.14 \metricgainpos{+0.79} & 7.48 \metricgainpos{+1.51} & 8.16 \metricgainpos{+0.47} & 8.39 \metricgainpos{+1.74} & 10.09 \metricgainpos{+1.23} & 10.95 \metricgainpos{+2.29} & 8.05 \metricgainpos{+0.31} & 8.42 \metricgainpos{+0.94} \\
Llama & \hybridbest{\textbf{9.03}} \metricgainpos{\hybridbest{\textbf{+1.22}}} & \hybridbest{\textbf{9.35}} \metricgainpos{\hybridbest{+1.43}} & \hybridbest{\textbf{10.33}} \metricgainpos{\hybridbest{\textbf{+0.71}}} & \hybridbest{\textbf{10.67}} \metricgainpos{\hybridbest{\textbf{+2.00}}} & \hybridbest{12.28} \metricgainpos{\hybridbest{\textbf{+1.70}}} & \hybridbest{12.58} \metricgainpos{\hybridbest{\textbf{+3.43}}} & \hybridbest{\textbf{9.70}} \metricgainpos{\hybridbest{\textbf{+0.66}}} & \hybridbest{\textbf{9.69}} \metricgainpos{\hybridbest{+1.00}} \\
Qwen3 & 8.05 \metricgainpos{+0.23} & \hybridsecond{8.58} \metricgainpos{\hybridsecond{+0.99}} & \hybridsecond{9.76} \metricgainpos{\hybridsecond{+0.58}} & 9.88 \metricgainpos{+0.96} & \hybridsecond{11.51} \metricgainpos{\hybridsecond{+0.61}} & \hybridsecond{11.54} \metricgainpos{\hybridsecond{+2.33}} & 8.63 \metricloss{-0.34} & 8.34 \metricgainpos{+1.15} \\
Qwen3-VL & \hybridsecond{8.51} \metricgainpos{\hybridsecond{+0.53}} & 8.25 \metricgainpos{+1.28} & 9.68 \metricgainpos{+0.26} & 9.93 \metricgainpos{+1.22} & 10.44 \metricgainpos{+0.96} & 11.01 \metricgainpos{+1.33} & \hybridsecond{9.21} \metricgainpos{\hybridsecond{+0.52}} & 8.47 \metricgainpos{+1.03} \\
Qwen3-VL-Photo & 8.84 \metricloss{-0.54} & 8.44 \metricgainpos{+1.02} & 10.19 \metricloss{-0.38} & \hybridsecond{10.08} \metricgainpos{\hybridsecond{+1.17}} & 10.53 \metricgainpos{+0.48} & 10.82 \metricgainpos{+2.22} & 9.29 \metricloss{-0.09} & \hybridsecond{8.66} \metricgainpos{\hybridsecond{+1.06}} \\
\midrule
ColBERT & 8.39 \metricgainpos{+0.24} & 8.03 \metricgainpos{+0.32} & 8.05 \metricloss{-0.27} & 7.85 \metricloss{-0.23} & 11.66 \metricgainpos{+0.14} & 11.59 \metricgainpos{+0.33} & 8.71 \metricgainpos{+0.04} & 8.67 \metricgainpos{+0.11} \\
ColPali & 0.33 & -- & 0.02 & -- & 0.00 & -- & 0.49 & -- \\
\bottomrule
\end{tabular}
}
\end{table}

\noindent\textbf{Observation 8. Retrieval improvements only partially transfer to RAG record selection.}
Table~\ref{tab:rag_result} evaluates whether retrieved records can support downstream record selection by generators.
Hybrid retrieval often improves Macro-F1, with the largest gain being +3.43 for Gemma4-31B with Llama w/ R.
However, the absolute scores remain low: the strongest raw score is 13.36, and even the best hybrid total remains far below the retrieval Recall@10 in Table~\ref{tab:main_result}.
Thus, \ourDataset{} also exposes a downstream bottleneck: after retrieval, generators still struggle to select the correct verified record set from the returned records.
% \section{Conclusion}

% \ourDataset{} benchmarks record retrieval over five heterogeneous sources, including relational tables, textual documents, image repositories, spatial databases, and knowledge graphs.
% %
% It consists of 857 QA pairs instantiated from Yelp business records.
% %
% We build \ourDataset{} in an answer-driven manner by first selecting some candidate answer records and then generating the corresponding natural-language questions.
% %
% We have further extensively evaluated it via five families of retrievers under the same evaluation metrics.
% %
% The results show our benchmark is effective for evaluating record retrieval on heterogeneous sources, yet existing methods have much room for accuracy improvement. 
% %
% In the future, we will study more capable methods for retrieving records from heterogeneous sources.

\section{Conclusion}

\ourDataset{} benchmarks record retrieval for QA over five heterogeneous sources: relational tables, text documents, image repositories, spatial databases, and KGs.
It contains 857 QA pairs instantiated from Yelp business records, where each question combines two or three sources.
The answer-driven construction initializes relational-field constraints, expands candidates through missing-value recovery, instantiates source-specific constraints, and applies contradiction detection plus human validation before finalizing the dataset.
The experiments show that \ourDataset{} remains challenging for current retrieval methods.
The best setting, hybrid retrieval with Llama and reranking, reaches only 32.78 Recall@10, while Self-RAG obtains the best MRR@10 at 25.26 but does not achieve the best Recall@10.
Source-combination and RAG analyses analyses further show that current methods struggle with spatial-KG cases, source-specific constraint composition, and downstream record selection.
These findings leave substantial room for methods that can retrieve and verify records across heterogeneous sources.
In the future, we will study more capable methods for retrieving records from heterogeneous sources.

\section{Limitations}
\label{limitation}

\ourDataset{} is instantiated from Yelp business records, so its current scope is local business search rather than all record-centric applications.
Although the construction workflow can be extended by adding new source-specific constraints $h$ and support sets $\mathcal{V}_h$, new domains or source fields still require source-specific preprocessing and verification rules.
Finally, construction relies on LLM/VLM judgments, threshold-based filtering, and manual checks for verification and contradiction detection, which introduces annotation cost, token cost, and possible residual noise.

\bibliographystyle{plainnat}
\bibliography{references}

\appendix

\section{KG Construction Details}
\label{app:kg_construction}

This section describes the feature-centric graph $\mathcal{G}$ used for KG source fields in \ourDataset{}.

Yelp reviews describe subjective business features rather than clean encyclopedic facts. We therefore construct a sentiment-aware heterogeneous graph $\mathcal{G}=(\mathcal{V},\mathcal{E})$ with three node types: users ($U$), businesses ($B$), and extracted features ($F$), such as ``quiet ambiance'' or ``friendly staff.'' The graph includes review edges $U\to B$ and polarity-aware feature edges $U\to F$ and $B\to F$, which store confidence scores and allow KG constraints to capture review-implied requirements that are absent from relational fields. We extract typed \textit{(feature, polarity, confidence)} tuples from about \textbf{900M} raw review tokens with GPT-OSS-120B and a seed ontology.
Because LLM extraction can split one concept into surface variants (e.g., ``24-hour service'' and ``open 24 hours''), we canonicalize synonymous and near-duplicate feature strings before graph construction. This reduces the feature inventory from 184,009 raw strings to 53,832 canonical features.

\begin{table}[h]
\centering
\caption{Statistics of the KG source graph $\mathcal{G}$.}
\label{tab:graph_stats}
\renewcommand{\arraystretch}{1.02}
\setlength{\tabcolsep}{6pt}
\footnotesize
\begin{tabular}{lrlr}
\toprule
Metric & Value & Metric & Value \\
\midrule
Nodes $|\mathcal{V}|$ & 237,082 & Edges $|\mathcal{E}|$ & 4,011,327 \\
Users $U$ & 80,315 & Avg. degree $U$ & 31.88 \\
Businesses $B$ & 102,935 & Avg. degree $B$ & 24.43 \\
Features $F$ & 53,832 & Avg. degree $F$ & 54.76 \\
Edge density & $1.43{\times}10^{-4}$ & Raw $\to$ final features & $184,009 \to 53,832$ \\
Redundancy score & $0.914 \to 0.862$ & Feature reduction & 70.7\% \\
\bottomrule
\end{tabular}
\end{table}

\section{Construction Models and Settings}
\label{app:construction_settings}

Table~\ref{tab:construction_models} summarizes the model roles and thresholds used during dataset construction. These models are separate from the retrieval baselines evaluated in Section~\ref{sec:experiments}.

\begin{table*}[t]
\centering
\caption{Construction-stage models and key settings.}
\label{tab:construction_models}
\renewcommand{\arraystretch}{1.05}
\setlength{\tabcolsep}{4pt}
\footnotesize
\begin{tabular}{p{0.25\textwidth}p{0.33\textwidth}p{0.34\textwidth}}
\toprule
Component & Model or setting & Use \\
\midrule
Source-specific constraint generation
& \texttt{gpt-5}
& Instantiates source-specific constraints $\mathcal{H}$ from other source fields. \\
Question verbalization
& \texttt{gpt-5}
& Verbalizes the final natural-language question $q$. \\
Support-set audit
& \texttt{gpt-5}
& Checks support sets $\mathcal{V}_h$ for constraints in $\mathcal{H}$. \\
Contradiction audit
& \texttt{gpt-5}
& Detects text/image contradictions against $q$. \\
Text embedding
& \texttt{qwen3-embedding-8B}
& Searches serialized fields, reviews, and tips. \\
Photo embedding
& \texttt{doubao-embedding-vision-250615}$^\dagger$
& Searches business photos. \\
Text reranking
& \texttt{Qwen3-Reranker-8B}
& Reranks text matches before verification. \\
Photo reranking
& \texttt{Qwen3-VL-Reranker-8B}
& Reranks photo matches before verification. \\
\midrule
Recovery/search depth
& \texttt{top\_k=300}
& Missing-value recovery and support-set checks. \\
Judge threshold
& \texttt{llm\_judge\_threshold=0.7}
& Filters LLM/VL-judge support judgments by confidence. \\
Text coarse threshold
& \texttt{text\_coarse\_thres=0.65}
& Coarse review/tip matches. \\
Text rerank threshold
& \texttt{text\_rerank\_thres=0.6}
& Reranked review/tip matches. \\
Photo coarse threshold
& \texttt{image\_coarse\_thres=0.65}
& Coarse photo matches. \\
Photo rerank threshold
& \texttt{image\_reranker\_thres=0.25}
& Reranked photo matches. \\
Contradiction threshold
& \texttt{contradiction\_ratio\_thres=0.15}
& Text/image contradiction audit. \\
\bottomrule
\end{tabular}

\vspace{1pt}
\begin{minipage}{0.98\textwidth}
\scriptsize
$^\dagger$ We used this closed-source photo embedding model during dataset construction because it gave the strongest photo retrieval performance in our construction-time trials.
\end{minipage}
\end{table*}

\subsection{Relational-Field Sampling Tiers}
\label{app:field_tiers}

Table~\ref{tab:field_tiers} lists the relational-field tiers used to form $F$ during relational initialization.

\begin{table*}[t]
\centering
\caption{Relational-field tiers used to initialize $F$.}
\label{tab:field_tiers}
\renewcommand{\arraystretch}{1.05}
\setlength{\tabcolsep}{4pt}
\scriptsize
\begin{tabular}{p{0.15\textwidth}p{0.07\textwidth}p{0.72\textwidth}}
\toprule
Tier & Count & Fields \\
\midrule
\textit{Core}
& 5
& \texttt{categories}, \texttt{city}, \texttt{state}, \texttt{stars}, \texttt{is\_open} \\
\midrule
\textit{High-value} & 10
& \texttt{business\_accepts\_credit\_cards}, \texttt{restaurants\_take\_out}, \texttt{restaurants\_delivery}, \texttt{good\_for\_kids}, \texttt{wheelchair\_accessible}, \texttt{restaurants\_price\_range2}, \texttt{outdoor\_seating}, \texttt{wifi}, \texttt{alcohol}, \texttt{restaurants\_reservations} \\
\midrule
\textit{Supplementary}
& 64
& \texttt{restaurants\_counter\_service}, \texttt{open\_24\_hours}, \texttt{ambience\_casual}, \texttt{has\_tv}, \texttt{restaurants\_good\_for\_groups}, \texttt{bike\_parking}, \texttt{dogs\_allowed}, \texttt{drive\_thru}, \texttt{caters}, \texttt{ambience\_romantic}, \texttt{by\_appointment\_only}, \texttt{happy\_hour}, \texttt{restaurants\_attire}, \texttt{noise\_level}, \texttt{restaurants\_table\_service}, \texttt{ambience\_classy}, \texttt{ambience\_trendy}, \texttt{music\_background\_music}, \texttt{ambience\_intimate}, \texttt{ambience\_touristy}, \texttt{ambience\_hipster}, \texttt{ambience\_divey}, \texttt{ambience\_upscale}, \texttt{music\_live}, \texttt{music\_dj}, \texttt{smoking}, \texttt{hair\_specializes\_in\_coloring}, \texttt{coat\_check}, \texttt{good\_for\_dancing}, \texttt{corkage}, \texttt{byob}, \texttt{business\_accepts\_bitcoin}, \texttt{accepts\_insurance}, \texttt{byob\_corkage}, \texttt{ages\_allowed}, \texttt{music\_jukebox}, \texttt{music\_karaoke}, \texttt{music\_no\_music}, \texttt{music\_video}, \texttt{hair\_specializes\_in\_extensions}, \texttt{hair\_specializes\_in\_kids}, \texttt{hair\_specializes\_in\_perms}, \texttt{hair\_specializes\_in\_straightperms}, \texttt{hair\_specializes\_in\_africanamerican}, \texttt{hair\_specializes\_in\_asian}, \texttt{hair\_specializes\_in\_curly}, \texttt{best\_nights\_friday}, \texttt{best\_nights\_saturday}, \texttt{best\_nights\_thursday}, \texttt{best\_nights\_wednesday}, \texttt{best\_nights\_tuesday}, \texttt{best\_nights\_sunday}, \texttt{best\_nights\_monday}, \texttt{business\_parking\_garage}, \texttt{business\_parking\_lot}, \texttt{business\_parking\_street}, \texttt{business\_parking\_valet}, \texttt{business\_parking\_validated}, \texttt{good\_for\_meal\_breakfast}, \texttt{good\_for\_meal\_brunch}, \texttt{good\_for\_meal\_lunch}, \texttt{good\_for\_meal\_dinner}, \texttt{good\_for\_meal\_dessert}, \texttt{good\_for\_meal\_latenight} \\
\bottomrule
\end{tabular}
\end{table*}

\subsection{Contradiction Detection and Answer-Set Ranking}
\label{app:construction_audit_ranking}

After support-set filtering, the construction auditor removes records whose source bundle contradicts the final question $q$.
For each retained record $r$, we retrieve question-relevant reviews/tips and photos using the embedding of $q$.
The prompts in Figures~\ref{fig:text_contradiction_prompt} and~\ref{fig:visual_contradiction_prompt} label each retrieved item as \textit{compatible} or \textit{contradictory}; neutral, irrelevant, or ambiguous items are treated as compatible unless they explicitly negate a requirement in $q$.
Let $\rho_{\mathcal{T}}(r)$ and $\rho_{\mathcal{I}}(r)$ be the contradiction ratios over the retrieved text and image items.
A record is removed if either ratio exceeds the contradiction threshold $\tau_c=0.15$ in Table~\ref{tab:construction_models}.
Surviving records are ranked using the verification confidences produced during construction.
Let $\mathcal{J}(r)$ be the available verification channels for record $r$, including spatial, KG, text, image, and cross-source checks.
The cross-source checks verify text-derived constraints with photos and image-derived constraints with reviews/tips.
Given the confidence score $c_j(r)$ from channel $j$, we compute
\begin{equation}
s(r)=\sum_{j\in\mathcal{J}(r)} c_j(r)-\rho_{\mathcal{T}}(r)-\rho_{\mathcal{I}}(r).
\end{equation}
A higher score means that a record has stronger source support and fewer detected contradictions, so it should be ranked higher.
Records in the verified answer set are ranked by $s(r)$ in descending order.

\begin{figure}[H]
\begin{AIbox}{Prompt for Text Contradiction Audit}
You are a text contradiction auditor.
Given a user question and a review/tip segment, decide whether the segment explicitly contradicts the question.

\textbf{Labels:}
\begin{itemize}[leftmargin=*, noitemsep, topsep=2pt]
    \item \texttt{CONTRADICTION}: the segment explicitly negates the question requirement, e.g., ``good service'' vs. ``rudest staff ever.''
    \item \texttt{COMPATIBLE}: the segment supports, is neutral to, or is irrelevant to the question.
\end{itemize}

\textbf{Input:}\\
User Question: \{question\}\\
Review/Tip Segment: \{review\}

\textbf{Output:}\\
Return JSON only:
\texttt{\{"label": "COMPATIBLE" or "CONTRADICTION", "reason": "brief explanation"\}}.
\end{AIbox}
\caption{The text contradiction audit prompt.}
\label{fig:text_contradiction_prompt}
\end{figure}

\begin{figure}[H]
\begin{AIbox}{Prompt for Visual Contradiction Audit}
You are a visual contradiction auditor.
Given a user question and a photo, decide whether the photo explicitly contradicts the question.

\textbf{Labels:}
\begin{itemize}[leftmargin=*, noitemsep, topsep=2pt]
    \item \texttt{CONTRADICTION}: the photo clearly negates the question requirement or shows an opposite environment.
    \item \texttt{COMPATIBLE}: the photo supports, is neutral to, or is irrelevant to the question.
\end{itemize}

\textbf{Input:}\\
User Question: \{question\}\\
Photo: \{photo\}

\textbf{Output:}\\
Return JSON only:
\texttt{\{"label": "COMPATIBLE" or "CONTRADICTION", "reason": "brief explanation"\}}.
\end{AIbox}
\caption{The visual contradiction audit prompt.}
\label{fig:visual_contradiction_prompt}
\end{figure}

\section{Human Validation and Query-Diversity Metrics}
\label{app:quality_validation}

We compute the query-diversity metrics in Table~\ref{tab:quality_audit} on \ourDataset{} following STARK~\cite{stark}.
Word Entropy measures lexical distributional diversity, defined as $H=-\sum_w p(w)\log_2 p(w)$ over the word distribution of questions.
Type-Token Ratio (TTR) measures lexical variety, defined as $\mathrm{TTR}=|V|/M$, where $V$ is the set of unique words and $M$ is the total number of words.

For human validation, we use the naturalness, diversity, and practicality dimensions from STARK~\cite{stark} as a protocol reference.
\ourDataset{} contains 857 questions across ten source-composition subsets.
We sample 200 questions with subset-proportional stratified sampling and ask 13 graduate annotators, including PhD and MPhil students, to rate the anonymized questions on a 1--5 Likert scale.
For each dimension, we compute subset-level rates and aggregate them with subset weights.
The \textit{positive rate} uses score $\ge 4$, and the \textit{non-negative rate} uses score $\ge 3$.
Table~\ref{tab:quality_audit} reports the resulting query-diversity and human-validation results.

\section{Evaluation-Time Retrieval Implementation}
\label{app:retrieval_algorithm}

Algorithm~\ref{alg:retrieval_system} summarizes the retrieval implementation.
Each method retrieves source-level content, maps it back to records in $\mathcal{R}$, and scores the final ranked records against $\mathcal{V}_{\mathcal{H}}^\star$.

\begin{algorithm*}[t]
\small
\caption{Evaluation-time retrieval and fusion.}
\label{alg:retrieval_system}
\KwIn{question $q$, configuration $c$}
\KwOut{top-$k$ record ranking $R$ and runtime summary $M$}
$k \leftarrow c.\texttt{top\_k}$; $k_b \leftarrow k \times c.\texttt{rerank\_recall\_multiplier}$\;
\If{$c.\texttt{pipeline\_type} = \texttt{colpali}$}{
  $V \leftarrow$ retrieve rendered-page hits at depth $k_b$\;
  \Return{top-$k$ records after record-level collapse of $V$}\;
}
\ForEach{active branch $b$}{
  \ForEach{source field $t \in c.\texttt{search\_targets}$}{
    $U_{b,t} \leftarrow$ retrieve top-$k_b$ source hits for $(q,b,t)$ and collapse them to records\;
    \If{$c.\texttt{enable\_reranker}$}{$U_{b,t} \leftarrow$ rerank $U_{b,t}$\;}
  }
  $R_b \leftarrow$ RRF fuse $\{U_{b,t}\}_t$\;
}
$R \leftarrow R_b$ if one branch is active; otherwise RRF fuse the active branch rankings\;
\Return{top-$k$ records in $R$ and runtime summary $M$}\;
\end{algorithm*}

\begin{table*}[t]
\centering
\caption{Evaluation-time retrieval settings.}
\label{tab:retrieval_config}
\renewcommand{\arraystretch}{1.08}
\setlength{\tabcolsep}{5pt}
\footnotesize
\begin{tabular}{p{0.22\textwidth}p{0.72\textwidth}}
\toprule
Control & Policy \\
\midrule
Source fields & Sparse, dense, and hybrid retrievers search serialized relational fields, Yelp reviews/tips, and business photos. \\
Vector backend & Vector-based retrievers use OceanBase HNSW indexes over three source stores: about 7.89M text entries, 0.20M photo entries, and 0.15M serialized relational-field entries. The text-store HNSW configuration uses cosine distance with $M=31$; OceanBase estimates 76.4GB for HNSW\_SQ and 51.8GB for HNSW\_BQ, with 7.8GB serving memory for HNSW\_BQ. \\
Depth & The final ranking depth is top-$k=10$; branch recall depth is $3k$ before record-level fusion. \\
Fusion & Sparse, dense, and hybrid branches use reciprocal rank fusion with $k=60$ after source hits are collapsed to records. \\
Reranking & w/ R reranks record-level candidates with Qwen3-VL-Reranker-8B before branch fusion. \\
Record-level collapse & Review snippets, captions, photos, and page hits from the same Yelp business are collapsed to one record before scoring. \\
Latency & Latency is the average end-to-end question time. \\
\bottomrule
\end{tabular}
\end{table*}

\begin{figure*}[t]
  \centering
  \fbox{\includegraphics[page=1,width=0.46\textwidth,height=0.42\textheight,keepaspectratio]{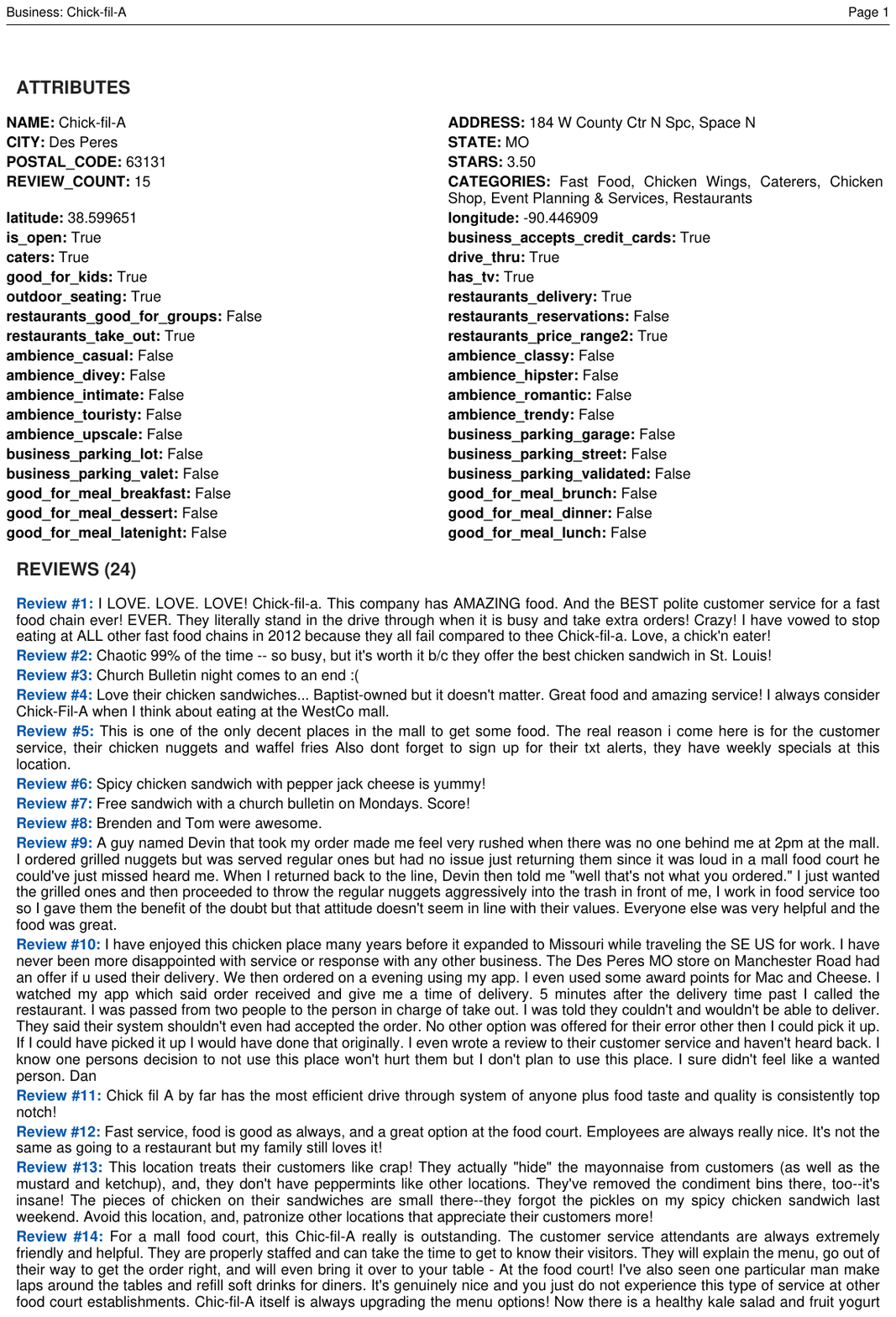}}
  \hfill
  \fbox{\includegraphics[page=2,width=0.46\textwidth,height=0.42\textheight,keepaspectratio]{sections/figures/colpali_input_example.pdf}}
  \caption{Example ColPali input rendered from one Yelp record. Retrieved pages are collapsed back to records for evaluation.}
  \label{fig:colpali_input_example}
\end{figure*}

\paragraph{ColPali input.}
For ColPali, each Yelp record is rendered into PDF pages containing serialized relational fields, reviews/tips, and photo thumbnails.
ColPali retrieves page-level hits, which are collapsed back to records before scoring.
Figure~\ref{fig:colpali_input_example} shows the rendered input.

\subsection{Agentic Retrieval}
\label{app:agentic_methods}

\textit{ReAct.}
We instantiate ReAct by interleaving reasoning and tool actions~\cite{react}.
At each step, the model observes the current retrieval state, selects one tool action, and receives the updated record list and source-field content.

\textit{Self-RAG.}
We implement Self-RAG as an inference-time reflection policy inspired by Self-RAG~\cite{selfrag}, without task-specific retraining.
The model emits a reflection label to decide whether to retrieve more records, reviews, or photos, critique current source-field content, or finish.

Both agentic methods use the same tool inventory in Table~\ref{tab:agentic_tools}; they differ only in the policy for selecting the next tool.

\begin{table}[H]
\centering
\caption{Tool inventory for agentic retrieval.}
\label{tab:agentic_tools}
\renewcommand{\arraystretch}{1.08}
\setlength{\tabcolsep}{4pt}
\footnotesize
\begin{tabular}{p{0.30\columnwidth}p{0.62\columnwidth}}
\toprule
Tool & Role in the retrieval workflow \\
\midrule
\texttt{Search\_Business} & Retrieves records from serialized relational fields and profile metadata with Qwen3-Embedding-8B. \\
\texttt{Geo\_Filter} & Applies spatial constraints, including radius search, directional relations, and filtering around an anchor business. \\
\texttt{Search\_Reviews} & Retrieves reviews/tips with Qwen3-Embedding-8B for text source constraints. \\
\texttt{Search\_Photos} & Retrieves business photos with Qwen3-VL-Embedding-8B for image source constraints. \\
\texttt{Finish} & Returns the final top-10 record ranking. \\
\bottomrule
\end{tabular}
\end{table}

\section{RAG Experiment Details}
\label{app:rag_details}

The RAG experiment evaluates downstream record selection. For each question, the generator receives $q$, an ordered top-30 record list after record-level collapse, and the available source content for each record. Reviews/tips are provided as text blocks, and photos are supplied as image inputs when available. The input is therefore a record-level list, not the final top-10 retrieval output.

The generator returns business names in JSON because target records in the Yelp instantiation are businesses. We resolve those names to unique business identifiers within the returned record list and score them against the verified answer set. Let $P_i$ be the predicted business-id set and $G_i$ the verified business-id set for question $i$. The per-question set-F1 is
\begin{equation}
\mathrm{F1}_i=\frac{2|P_i\cap G_i|}{|P_i|+|G_i|}.
\end{equation}
Macro-F1 averages $\mathrm{F1}_i$ over the 857 questions. Exact-set-match accuracy assigns 1 only when $P_i=G_i$ as unordered sets. Generation errors are counted in the 857-question denominator.

The prompt template used for the RAG record-selection task is listed with the other prompt templates in Appendix~\ref{app:examples}.

\section{Prompt Templates}
\label{app:examples}

The following prompt templates support RAG evaluation and construction reproducibility.

\begin{AIbox}{RAG record-selection prompt for the Yelp instantiation}
\footnotesize
\textbf{System.}
You are given a user question and an ordered list of businesses with source content. Select only the businesses that satisfy the question based on the provided source content. Do not invent businesses. Return at most 10 businesses. If more than 10 businesses appear to satisfy the question, keep only the best 10. Return JSON only in this exact format: \texttt{\{"selected\_businesses": ["business name 1", "business name 2"]\}}. If none of the businesses satisfy the question, return \texttt{\{"selected\_businesses": []\}}.

\medskip
\textbf{User.}
\texttt{User question: <question>}\\[2pt]
\texttt{Businesses are listed below in implicit ranking order. Use only the provided source content when selecting businesses.}\\[2pt]
\texttt{Business: <business\_name>}\\
\texttt{Source content 1: <text content>}\\[2pt]
\texttt{Business: <business\_name>}\\
\texttt{Source content 1: <photo content>}\\[2pt]
\texttt{Business: <business\_name>}\\
\texttt{Source content 1: <text content>}\\
\texttt{Source content 2: <text or photo content>}
\end{AIbox}

\begin{figure}[H]
\begin{AIbox}{Prompt for Photo Relevance Judgment}
You are a helpful photo-evidence assistant. Your task is to determine whether the information visible in the provided photo can be used to answer the given text query.

\textbf{Input:}
\begin{itemize}[leftmargin=*, noitemsep, topsep=2pt]
    \item A text query (question or request from the user).
    \item A photo.
\end{itemize}

\textbf{Your job:}
\begin{enumerate}[leftmargin=*, noitemsep, topsep=2pt]
    \item Carefully analyze the photo and the query.
    \item Decide if the photo evidence is relevant and can potentially be used to answer the query.
    \item Output a confidence score between 0 and 1 (float), where:
    \begin{itemize}[leftmargin=1em, noitemsep]
        \item 0 means the photo is completely irrelevant and cannot help answer the query.
        \item 1 means the photo is highly relevant and likely contains information that can directly answer the query.
    \end{itemize}
    \item The output MUST be in JSON format with the following structure: \\
    \texttt{\{ \\
    \hspace*{1em} "can\_answer": boolean, \ \ // True if the photo can potentially help answer the query, False otherwise \\
    \hspace*{1em} "confidence": float \ \ \ \ \ // A value between 0.0 and 1.0 indicating how likely the photo is useful for answering the query \\
    \}}
\end{enumerate}
Do not include any extra commentary or explanation. Just output the JSON.

\textbf{Input:} \\
Text\_query: \{query\}
\end{AIbox}
\caption{The photo relevance judgment prompt used to filter uninformative photos.}
\label{fig:vlm_judge_prompt}
\end{figure}

\begin{figure}[H]
\begin{AIbox}{Prompt for Semantic Relevance Judgment}
You are an expert semantic relevance analyzer specialized in business reviews and user queries. \\
Your task is to determine whether a given user review is relevant to a specific query. \\
Relevance is defined as the review containing information that directly addresses the intent, context, or key entities (e.g., products, services, or record fields) mentioned in the query.

\textbf{Guidelines:}
\begin{enumerate}[leftmargin=*, noitemsep, topsep=2pt]
    \item Focus on semantic meaning rather than keyword matching.
    \item Ignore irrelevant details in the review (e.g., general descriptions unrelated to the query).
    \item Consider the query's intent: recommendations, complaints, comparisons, or factual inquiries.
    \item Output a JSON object containing:
    \begin{itemize}[leftmargin=1em, noitemsep]
        \item \texttt{"judgement"}: \texttt{"Yes"} or \texttt{"No"} indicating relevance
        \item \texttt{"confidence"}: a decimal score between 0 and 1 representing your confidence in the judgement
    \end{itemize}
    \item Confidence score should reflect:
    \begin{itemize}[leftmargin=1em, noitemsep]
        \item 0.9-1.0: Clear and direct relevance/irrelevance
        \item 0.7-0.89: Strong evidence but some ambiguity
        \item 0.5-0.69: Moderate relevance with significant ambiguity
        \item Below 0.5: Weak or unclear relevance
    \end{itemize}
    \item Output only the JSON object with no additional explanations, headers, or text.
\end{enumerate}

\textbf{Example-1:} \\
Input-Review: The hotel's pool was closed for maintenance, and the staff didn't inform us at check-in. The room was clean but the disappointment ruined our stay. \\
Input-Query: Find hotels with well-maintained pools and responsive staff. \\
Output: \texttt{\{ "judgement": "NO", "confidence": 0.95 \}}

\textbf{Example-2:} \\
Input-Review: The delivery was fast, and the packaging was eco-friendly. The product itself worked as described, but the instructions were unclear. \\
Input-Query: Recommend brands with sustainable packaging practices. \\
Output: \texttt{\{ "judgement": "YES", "confidence": 0.85 \}}

\textbf{Real Case:} \\
Input-Review: \{input\_review\} \\
Input-Query: \{input\_query\} \\
Output:
\end{AIbox}
\caption{The semantic relevance judgment prompt used to filter irrelevant reviews.}
\label{fig:relevance_prompt}
\end{figure}

\begin{figure}[H]
\begin{AIbox}{Prompt for Feature Extraction}
\footnotesize
\textbf{\# Feature and Relationship Extraction (Single Review, Strict)}

\textbf{\#\# Task} \\
You will process EXACTLY ONE review or tip for ONE business. \\
Your job is to extract meaningful feature-level signals that reflect:
\begin{itemize}[noitemsep,topsep=0pt,leftmargin=*]
    \item what the USER cares about or reacts to
    \item how the BUSINESS performs on those features
\end{itemize}

\textbf{\#\# Canonical Feature List (Strongly Preferred)} \\
\texttt{[quiet\_ambiance, lively\_atmosphere, ambience\_romantic, cozy\_environment, modern\_design, ambience\_upscale, ambience\_casual, ambience\_intimate, spacious\_layout, crowded\_energy, ambience\_trendy, rustic\_charm, elegant\_style, industrial\_design, minimalist\_aesthetic, ambience\_upscale, peaceful\_environment, noisy\_energy, hip\_vibe, vintage\_style, artistic\_decor, historic\_charm, beachy\_vibe, mountain\_lodge\_feel, urban\_chic, suburban\_comfort, rooftop\_views, garden\_setting, library\_quiet, party\_energy, sophisticated\_mood, bohemian\_style, fast\_service, slow\_paced, friendly\_staff, knowledgeable\_service, attentive\_servers, efficient\_operations, personalized\_attention, professional\_service, accommodating\_staff, rude\_service, neglectful\_service, prompt\_service, leisurely\_service, disorganized\_service, well\_trained\_staff, multilingual\_staff, patient\_service, rushed\_service, customized\_experience, consistent\_service, inconsistent\_service, expert\_recommendations, spicy\_food, mild\_flavors, sweet\_dishes, savory\_options, umami\_rich, bitter\_notes, authentic\_flavors, fusion\_cuisine, traditional\_recipes, innovative\_dishes, comfort\_food, gourmet\_experience, homemade\_quality, bold\_flavors, delicate\_flavors, complex\_flavors, simple\_preparations, rich\_sauces, light\_preparations, smoky\_flavors, herbal\_notes, citrus\_infused, garlic\_forward, buttery\_textures, creamy\_sauces, crunchy\_textures, tender\_meats, flaky\_pastries, chewy\_textures, refreshing\_flavors, hearty\_portions, delicate\_presentation, dietary\_restrictions\_vegetarian, dietary\_restrictions\_vegan, dietary\_restrictions\_gluten\_free, dietary\_restrictions\_dairy\_free, keto\_friendly, low\_carb, organic\_ingredients, local\_produce, sustainable\_sourcing, long\_wait\_times, quick\_seating, reservation\_required, walk\_in\_friendly, affordable\_prices, expensive\_dining, good\_value, overpriced\_items, wifi, outdoor\_seating, wheelchair\_accessible, good\_for\_kids]}

You SHOULD prefer features from the list above. \\
You MAY introduce a new feature \textbf{only if}:
\begin{itemize}[noitemsep,topsep=0pt,leftmargin=*]
    \item the review clearly expresses an important concept
    \item no existing feature reasonably captures it
\end{itemize}
Do NOT invent vague or generic features.

\textbf{\#\# When to Return null} \\
If the text: is purely factual with no opinions; contains no clear preferences, complaints, or praise; or is too vague to support confident feature extraction. \\
THEN output: \texttt{null}. \\
Returning \texttt{null} means: you have successfully evaluated the text and explicitly decided there is no extractable feature-level signal. This is a VALID and EXPECTED outcome.

\textbf{\#\# Output Format (STRICT)} \\
- Output JSON ONLY. Do NOT wrap in a list. Do NOT include explanations or comments. \\
Either output: \\
\texttt{\{ \\
\hspace*{1em} "user": [ \{ "feature": "...", "sentiment": "positive|negative", "confidence": 0.0 \} ], \\
\hspace*{1em} "business": [ \{ "feature": "...", "quality": "good|poor", "confidence": 0.0 \} ] \\
\}} \\
OR: \texttt{null}

\textbf{\#\# Review Text} \\
\{review\}
\end{AIbox}
\caption{The strict feature extraction prompt utilized in the pipeline.}
\label{fig:extraction_prompt}
\end{figure}

% \clearpage
% \input{sections/checklist}

\end{document}